\documentclass[twocolumn]{aastex63}
\usepackage{amsmath}
\usepackage{color}
\def\beq{\begin{eqnarray}}
\def\eeq{\end{eqnarray}}

\begin{document}

\title{Black hole hyperaccretion in collapsars. III. GRB timescale}

\correspondingauthor{Tong Liu}
\email{tongliu@xmu.edu.cn}

\author[0000-0002-9130-2586]{Yun-Feng Wei}
\affiliation{Department of Astronomy, Xiamen University, Xiamen, Fujian 361005, China}

\author[0000-0001-8678-6291]{Tong Liu}
\affiliation{Department of Astronomy, Xiamen University, Xiamen, Fujian 361005, China}

\begin{abstract}
Gamma-ray bursts (GRBs) are classified into long and short populations (i.e., LGRBs and SGRBs) based on the observed bimodal distribution of duration $T_{90}$. Multimessenger observations indicated that most SGRBs and LGRBs should be powered by ultrarelativistic jets launched from black hole (BH) hyperaccretion in compact object mergers and massive collapsars, respectively. However, the duration criterion sometimes cannot correctly reflect the physical origin of a particular GRB. In the collapsar scenario, a GRB can be observed when the jet breaks out from the envelope and circumstellar medium successfully. The observed GRB duration reflects only the time that the engine operates after the jet breaks out. This work studies the propagation of jets driven by the neutrino annihilation or Blandford-Znajek mechanism in massive collapsars. The signatures of the progenitors for producing LGRBs, SGRBs, and failed GRBs in the collapsar scenario are exhibited. The competition between the mass supply onto the BH hyperaccretion and jet propagation into the envelope are definitely dependent on the density profiles of the collapsars. We show that duration and isotropic energy $E_{\rm{\gamma,iso}}$ of GRBs can help constrain the density profiles of collapsars.  Finally, we propose that a collapsar-origin SGRB, GRB 200826A, might originate from a neutrino-annihilation-dominated jet launched by a $\sim 10~M_\odot$ collapsar whose progenitor's envelope has been stripped.
\end{abstract}

\keywords{accretion, accretion disks - black hole physics - gamma-ray burst: general}

\section{Introduction}

Gamma-ray bursts (GRBs) are among the most luminous explosions in the universe. According to the observed bimodal distribution of duration $T_{90}$ (the time interval during which a detector observes $5\%$ to $95\%$ of the total gamma-ray fluence), GRBs are classified into long and short populations \citep{Kouveliotou1993}. Short-duration GRBs (SGRBs, $T_{90}<2 ~\rm{s}$) and long-duration GRBs (LGRBs, $T_{90}>2 ~\rm{s}$) are generally considered to be linked to binary neutron star (NS) or NS-black hole (BH) mergers \citep[e.g.,][]{Abbott2017} and deaths of massive stars \citep[e.g.,][]{Woosley2006}, respectively. However, the duration is sometimes not a reliable indicator of the GRB physical origin. At present, some LGRBs have the statistical properties of SGRBs and are suggested to come from compact object mergers \citep[e.g.,][]{Gal-Yam2006,Gehrels2006}. Meanwhile, some SGRBs have been suggested to originate from massive collapsars \citep[e.g.,][]{Zhang2007,Levesque2010,Thone2011,Xin2011}. Therefore, \citet{Zhang2006} and \citet{Zhang2009} suggested that GRBs should be classified physically into two classes: Type I (compact-object-merger origin) and Type II (massive-collapsar origin).

There are two ways that the duration of Type II GRBs may be less than $2 ~\rm{s}$. First, they can be naturally caused by the ``tip-of-iceberg" effect \citep[e.g.,][]{Lu2014}. In this case, a real LGRB may be observed as a ``short" one if the majority of gamma-ray emission episodes are below the detection threshold of GRB detectors. Second, massive stars indeed produce SGRBs. In the collapsar scenario \citep[e.g.,][]{Woosley1993,MacFadyen1999,Woosley2002,Zhang2004,Woosley2012}, the collapsed core of a massive star forms a fast-rotating BH surrounded by a hyperaccretion disk, which launches the relativistic jets along its rotation axis. If one of the jets can break out from the envelope and circumstellar medium in the prompt emission phase and is along the line of sight, then an observable GRB is triggered. Subsequently, the jet will be decelerated in the medium or winds to produce the multi-band afterglows. Thus, the observed GRB duration reflects only the duration of the central engine after the jet breaks out. Once the observed timescale is less than $2~\rm{s}$, this event is classified as an SGRB. If the jet fails to break out, then no GRB can be detected. Therefore, the duration of GRB mainly depends on jet propagation in the envelope.

Jet propagation in collapsars has been investigated in both analytical works \citep[e.g.,][]{Matzner2003,Bromberg2011a,Bromberg2012,Suwa2011,Matsumoto2015,Liu2018,Liu2019,Song2019} and numerical works \citep[e.g.,][]{Zhang2003,Mizuta2006,Mizuta2009,Mizuta2011,Morsony2007,Nagakura2011,Nagakura2012,Nagakura2013,Nakauchi2012,Tominaga2007}. These studies found that the propagation of jets in various types of massive stars, such as Wolf-Rayet (WR) stars and Population III (Pop III) stars, can produce GRBs. The detection of GRB 200826A confirms that collapsars can produce SGRBs, and this burst appears to sit on the brink between a successful and a failed GRB \citep[e.g.,][]{Ahumada2021,Zhang2021}. The purpose of this work is to systematically study the dependence of GRB durations on jet propagation in the collapsar scenario.

The paper is organized as follows. In Sections 2.1 and 2.2, we introduce the progenitor and jet models. In Section 2.3, we review the essential features of jet propagation and the corresponding characteristics of GRBs, such as the duration and energy of prompt emission. The results are shown in Section 3. In Section 4, we constrain the collapsar density profile of GRB 200826A. A brief summary is given in Section 5.

\section{Models}

\subsection{Progenitor model}

\begin{figure}
\centering
\includegraphics[angle=0,scale=0.33]{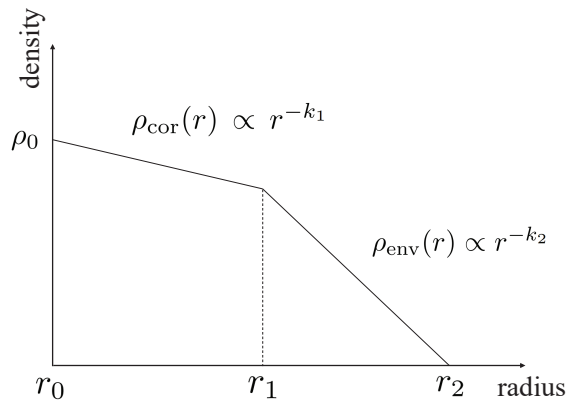}
\caption{The artificial density profile of collapsar. The density profiles are divided into two parts: star core and envelope. The density of the core is approximated as a power law with an index $-k_1$. The density of the envelope is approximated as a power law with an index $-k_2$. $r_1$ is the boundary between them. The outer boundary of collapsar $r_2$ is defined as the radius at which the density decreases to $10^{-10}~\rm{g}~\rm{cm}^{-3}$. $r_{\rm{0}}$ is the radius where the enclosed mass reaches $3 ~M_{\odot}$, and $\rho_0$ is the density at this radius.}
\end{figure}

In this paper, we construct a series of collapsar density profiles to investigate jet propagation. Note that these collapsars are evolved massive stars (from zero-age main sequence to the stage that the core begins to collapse) and have experienced mass loss. We assume that the precollapse star has a core-envelope structure \citep[see e.g.,][]{Kumar2008}. Then, the density profiles are divided into two parts: star core and envelope. As shown in Figure 1, the density profiles of the core and the envelope can be approximated as $\rho_{\rm{cor}} (r) \propto r^{-k_1}$ and $\rho_{\rm{env}} (r) \propto r^{-k_2}$, respectively. Note that $r_1$ is the boundary between the core and envelope. The outer boundary of collapsar $r_2$ is defined as the radius at which the density decreases to $10^{-10} ~\rm{g}~\rm{cm}^{-3}$. Actually, the exact stellar surface is difficult to calculate for the simulations of stellar evolution \citep[e.g.,][]{Suwa2011}. For different stellar models in \citet{Woosley2002}, the densities of the outermost layer of collapsars are different. Following the previous studies of jet propagation \citep[e.g.,][]{Nakauchi2012,Nakauchi2013}, we adjust them to the same value. Here, we adopt the point of $10^{-10} ~\rm{g}~\rm{cm}^{-3}$ as the effective collapsar surface in calculating the jet propagation. This density is low enough that the accretion of the envelope cannot ignite the central engine \citep[e.g.,][]{Liu2018}. $r_0$ is the radius where the enclosed mass reaches $3 ~M_{\odot}$, and $\rho _0$ is the density at this radius. Then the mass of collapsar can be calculated as:
\beq
\frac{M_{\rm{col}}}{M_{\odot }} = 3+ \int_{r_{0}}^{r_{1}}\rho_{\rm{cor}}4\pi r^{2} dr+\int_{r_{1}}^{r_{2}}\rho_{\rm{env}}4\pi r^{2} dr.
\eeq
Referring to the presupernova model \citep[e.g.,][]{Woosley2002,Woosley2007,Heger2010}, we adopt $r_0=5.5\times 10^{8}~\rm{cm}$ and $\rho_0=6\times 10^{6}~\rm{g}~\rm{cm}^{-3}$ as typical parameters in our calculations. By changing the values of $r_{\rm{1}}$ and $k_{\rm{2}}$, we can construct a series of density profiles. The minimum mass of the collapsar is set to $M_{\rm col}=10 ~M_{\odot}$, which corresponds to the typical mass of a carbon-oxygen WR star \citep{Matzner2003}. $r_{1}$ determines the size of the core. The core becomes larger as $r_{1}$ becomes larger. The value of $r_{\rm{1}}$ varies from $10^{9}$ to $10^{10}~ \rm{cm}$. $k_{1}$ and $k_{2}$ determine the structure of the core and envelope of the collapsar, respectively. The value of $k_{\rm{2}}$ varies from $5$ to $40$. When $k_{2}$ is small, the collapsar will have a thick envelope. The progenitor of such a collapsar might be a low metallicity massive star, which experiences little mass loss during evolution and thereby has a large envelope at the end of its life. Here the lower limit of $k_{2}$ is set to $5$. In fact, for some types of stars in their precollapse phase, $k_{2}$ is less than $5$. For example, the envelope of the red supergiant (RSG) is very thick ($\rho_{\rm{env}}\propto r^{-3/2}$). However, the jets cannot break out from these RSGs \citep[e.g.,][]{Suwa2011,Matsumoto2015}, which suggests that these RSGs are not the progenitors of GRBs. Therefore, we consider steeper envelope density profiles here. As discussed in Section 3, our results show that collapsars are unlikely to produce SGRBs when $k_{2}<5$. Thus, we set the lower limit of $k_{2}$ to $5$. When $k_{2}$ is very large, the collapsar almost loses its envelope. Such a collapsar might be a WR star, and jets can easily break out of them. According to all cases of the density profiles of collapsars, when $k_{2}>40$, the mass ratio of the envelope to collapsar is less than $1\%$, which suggests that the collapsar almost loses its envelope. Thus, we set the upper limit of $k_{2}$ to $40$.

Then, we calculate the mass supply rate of collapsars with different density profiles. The accretion timescale of matter at a radius $r$ is estimated as a free-fall timescale \citep[e.g.,][]{Woosley2012,Matsumoto2015}:
\beq
t_{\rm{ff}}=\sqrt{\frac{3\pi}{32G\bar{\rho}}}=\frac{\pi}{2}\sqrt{\frac{r^{3}}{2G M_r}},
\eeq
where $\bar{\rho}=3M_{r}/(4\pi r^{3})$ is the mean density within $r$ and $M_{r}$ is the mass coordinate. Then, the mass supply rate can be evaluated as \citep[e.g.,][]{Suwa2011}:
\beq
\dot{M}_{\rm{pro}}=\frac{dM_{r}}{dt_{\rm{ff}}}=\frac{dM_{r}/dr}{dt_{\rm{ff}}/dt}=\frac{2M_{r}}{t_{\rm{ff}}}(\frac{\rho}{\bar{\rho}-\rho }).
\eeq
where $\rho$ is the mass density of the collapsar. We roughly set the above mass supply rate equals to the accretion rate $\dot{M}$ of the BH hyperaccretion system in the center of the collapsar \citep[e.g.,][]{Kashiyama2013,Nakauchi2013}. We neglect the mass loss in the process. Due to the angular momentum redistribution, an outflow is launched when the matter falls onto the outer boundary of the disk. The accretion of rotating gas through the disk is significantly reduced compared with the accretion of nonrotating gas \citep[e.g.,][]{Igumenshchev2003,Proga2003}. Moreover, the outflows might be launched from the disk when the mass accretion rate is high \citep[e.g.,][]{Liu2018,Song2019}. As a result, only a fraction of supplied mass is eventually accreted into the BH and the accretion rate we calculated is overestimated. In other words, the larger $r_1$ or $k_2$ might be required in the below results if the outflows are considered.

\subsection{Jet model}

The relativistic jets are launched from a BH hyperaccretion system in the center of a massive collapsar. The mechanism for converting the accretion energy or BH rotation energy into directed relativistic outflow remains uncertain. There are two well-known candidate mechanisms: neutrino annihilation and magnetohydrodynamic mechanisms, including the Blandford-Znajek (BZ) process \citep{Blandford1977}.

For the neutrino annihilation, if the accretion rate is very high, then photons are trapped in the disk, and only generous neutrinos can escape from the disk surface and annihilate in the space out of the disk to produce relativistic electron-positron jets. This accretion disk is called a neutrino-dominated accretion flow (NDAF), which has been widely investigated in previous works \citep[e.g.,][]{Popham1999,Narayan2001,Kohri2002,Lee2005,Gu2006,Chen2007,Janiuk2007,Kawanaka2007,Liu2007,Liu2015a,Liu2015b,Liu2017,Lei2009,Xue2013,Song2016,Nagataki2018}. For the BZ mechanism, the BH rotation energy can be converted into the Poynting flux jet by a
surrounding magnetic field \citep[e.g.,][]{Lee2000a,Lee2000b,Mizuno2004,McKinney2004,Barkov2008,Nagataki2009,Lei2013,Lei2017,Liu2015a,Wu2013}. For the same BH spin parameter and accretion rate, the neutrino annihilation luminosity is approximately two orders of magnitude smaller than the BZ luminosity \citep[e.g.,][]{Liu2015a,Liu2017}. Here we adopt two jet models for jet producing mechanisms in this work. We assume that the jet is formed when the mass of BH reaches 3 $M_{\odot }$ since it has little effect on the luminosity \citep[e.g.,][]{Qu2022} and set $t=0$ at this time.

First, we assume that the jet is driven by neutrino annihilation. The jet luminosity is estimated as the neutrino annihilation luminosity: $L_{\rm{j}}=L_{\rm{\nu }\rm{\bar{\nu}}}$. By investigating one-dimensional global solutions of NDAFs in the Kerr metric, \citet{Xue2013} derived fitting formulae for the annihilation luminosity, which is written as
\beq
\log L_{{\nu }{\bar{\nu}}}({\rm erg\,s^{-1}})\approx 49.50+2.45a_{*}+2.17\log\dot{m},
\eeq
where $a_{*}$ is the mean dimensionless BH spin parameter, and $\dot{m}=\dot{M}/M_\odot ~\rm s^{-1}$ is the dimensionless accretion rate. Here, we adopt $a_{*}=0.9$ \citep[e.g.,][]{Song2020,Wei2021}.

Second, the jet is assumed to be driven by the BZ process and neutrino annihilation together. If the jet is produced by the BZ process, the required large-scale magnetic field may need to be maintained by the NDAF. These two mechanisms may coexist in a BH hyperaccretion system \citep[e.g.,][]{Liu2015a,Liu2017,Liu2018}. The neutrino annihilation luminosity mainly plays a role in the early stage of accretion. When the accretion rate decreases, the neutrino related process will be terminated and the BZ mechanism dominates the jet luminosity. Therefore, the duration of the central engine for the second model is obviously longer than that for the first model. For the BZ process, the jet luminosity is given by \citep[e.g.,][]{Komissarov2010,Suwa2011,Matsumoto2015}
\beq
L_{j}=\eta \dot{M}c^{2}.
\eeq
where $\eta$ is the efficiency parameter. The value of $\eta$ is still uncertain in the collapsar scenario. Here, we assume that $\eta =6.2\times 10^{-4}$, which is a typical value for a WR star progenitor model \citep[e.g.,][]{Suwa2011}. Then, the jet luminosity can be written as
\beq
L_{j}=1.10\times 10^{51}\dot{m}+10^{49.50+2.45a_{*}+2.17\log\dot{m}}.
\eeq
Here, the BH spin parameter is also adopted as $a_{*}=0.9$.

\begin{figure*}
\centering
\includegraphics[angle=0,scale=0.3]{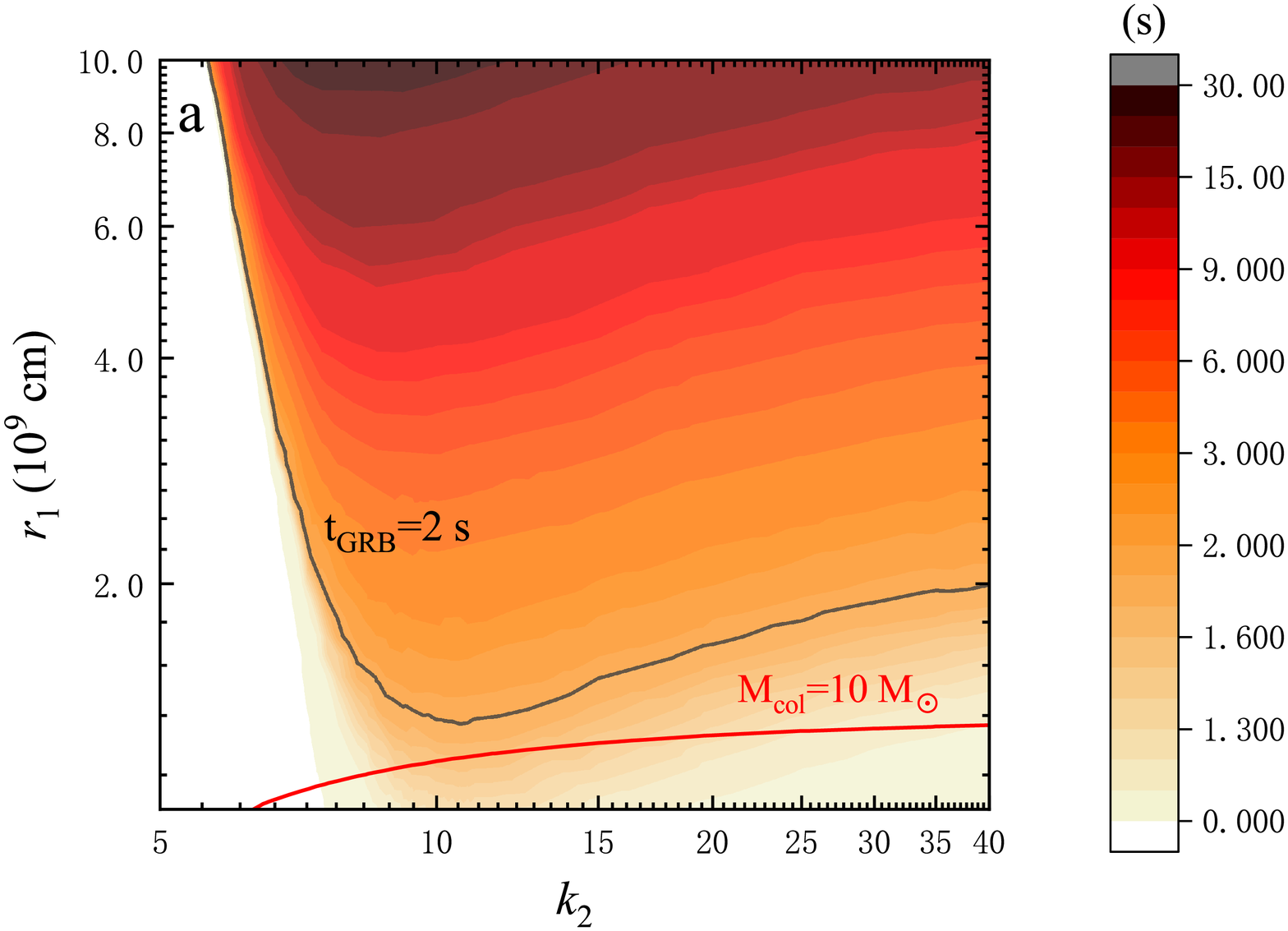}
\includegraphics[angle=0,scale=0.3]{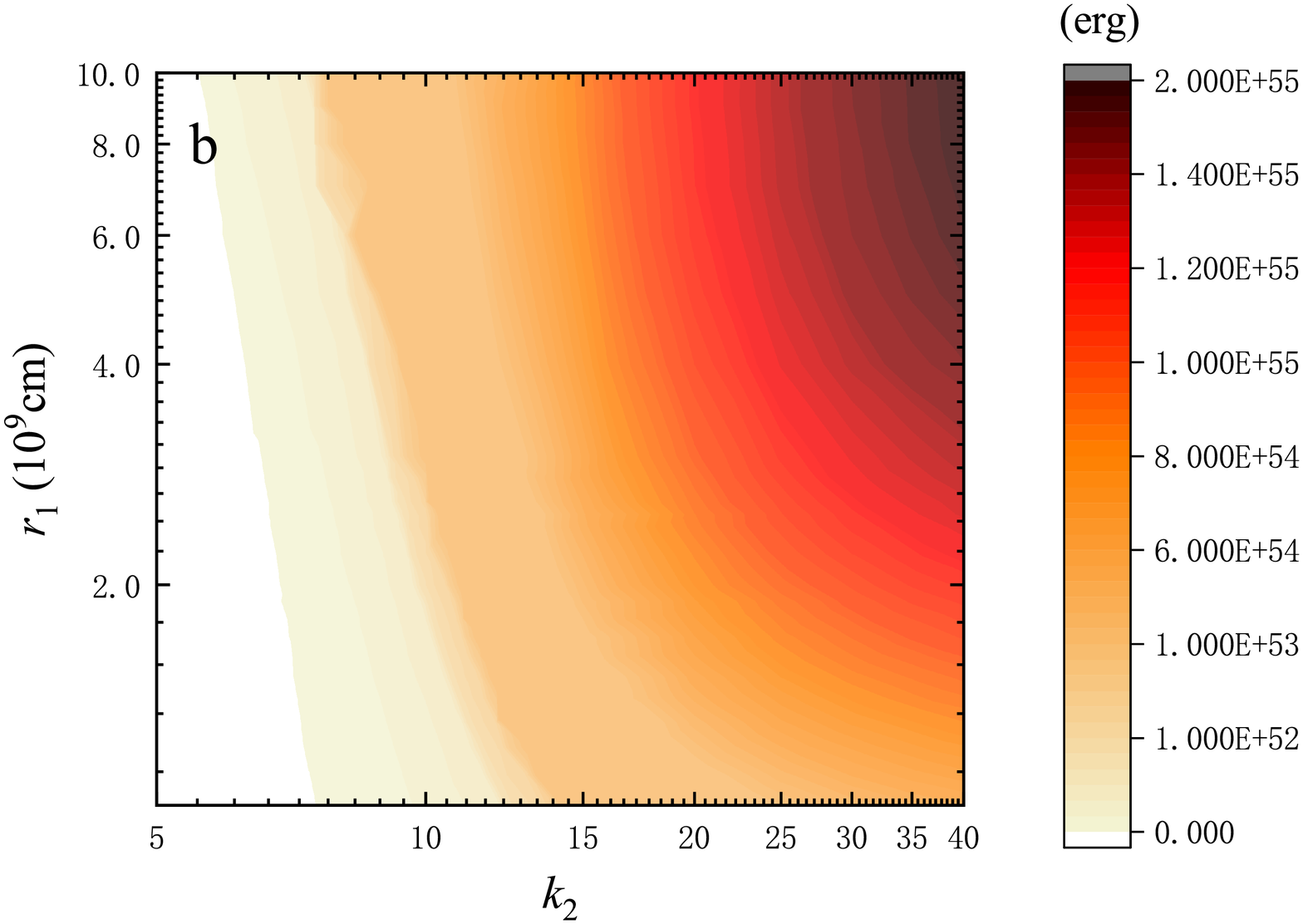}
\caption{Plane (a): GRB duration $t_{\rm{GRB}}$ as function of $r_{\rm{1}}$ and $k_{\rm{2}}$ for the first jet model. Black line corresponds to $t_{\rm{GRB}}=2~\rm{s}$. Red line corresponds to collapsar mass $M_{\rm col}=10 ~M_{\odot}$. Plane (b): isotropic energy $E_{\rm{\gamma,iso} }$ as function of $r_{\rm{1}}$ and $k_{\rm{2}}$.}
\end{figure*}

\subsection{Jet propagation}

The essential features of jet propagation in an envelope were described in previous works. After a relativistic jet is launched, it pushes the collapsar matter, and two shocks are formed at the jet head. One is the forward shock, which sweeps the collapsar matter. The other is the reverse shock, which would decelerate the head of the jet. The velocity of the jet head can be calculated from the pressure balance at the interface of the jet and the envelope as \citep[e.g.,][]{Matzner2003,Bromberg2011b,Matsumoto2015,Liu2018}
\beq
\beta _{h}=\frac{1}{1+\tilde{L}^{-1/2}},
\eeq
where
\beq
\tilde{L}\equiv \frac{L_{j}(t-r_{\rm{h}}/c)}{\pi r_{\rm{h}}^{2}\theta _{j}^{2}\rho (r_{\rm{h}})c^{3}}
\eeq
and $\theta _{j}$ is the jet half-opening angle. We adopt $\theta _{j}=5^{\circ}$ in our calculations. $r_{\rm h}(t)=\int_{0}^{t}c\beta_{\rm{h}} dt$ is the radius of the jet head. The jet breakout time $t_{b}$ is defined as $r_{\rm{h}}(t_{b})=r_{\rm{2}}$. Only after a jet breaks out from the collapsar can a GRB be observed. In the rest frame, the duration of GRB is $t_{\rm{GRB}}=t_{\rm{eng}}-t_{b}$, where $t_{\rm{eng}}$ is the duration of the central engine. We assume that the central engine operates until the jet luminosity decreases to $10^{47}~\rm{erg}~\rm{s}^{-1}$.

When the velocity of the jet head is nonrelativistic, the shocked material is pushed sideways to form a hot cocoon around the jet \citep{Matzner2003}. We assume that the jet energy goes through the shocked region into the cocoon before the jet reaches the collapsar surface. If the jet head can break out from the envelope successfully and the velocity of the jet head is larger than that of the cocoon, then the jet emission can be seen as a GRB \citep{Matzner2003,Toma2007}. The efficiency for converting the jet energy to radiation energy remains uncertain. Here following \citet{Nakauchi2012}, we assume the efficiency $\zeta$ is 10 $\%$. The energy of prompt emission can be estimated by
\beq
E_{\gamma}=\int_{t_{\rm{b}}}^{t_{\rm{eng}}}\zeta  L_{\rm{j}}(t)dt.
\eeq
Considering the beaming correction, the isotropic energy of prompt emission can be calculated by \citep[e.g.,][]{Yi2017}
\beq
E_{\gamma,\rm iso }=E_{\gamma}/f_{\rm{b}},
\eeq
where
\beq
f_{\rm{b}}=1-\rm{cos}\theta _{\rm{j}}\simeq \theta _{\rm{j}}^{2}/2.
\eeq

In this paper, we study the dependence of jet propagation on the collapsar density profile. Note that we consider the propagation of a jet in the stationary envelope in this work. The fallback process and jet propagation are separately calculated. Therefore, the feedback of the jet on the accretion is neglected. According to \citet{Nagakura2012}, some portion of the envelope ceases to fall due to jet propagation. However, a larger amount of matter can still be accreted into the BH. Thus, even considering the feedback of the jet, the jet may break out successfully and create a GRB. We also neglect the effect of the star rotation. \citet{Nagakura2013} investigated jet propagation through a rotating collapsing WR star. They found that the neutrino-driven jet can break out of the star by the different progenitor rotations and suggested that rapidly rotating stars are more likely to produce GRBs.

\section{Results}

\begin{figure*}
\centering
\includegraphics[angle=0,scale=0.31]{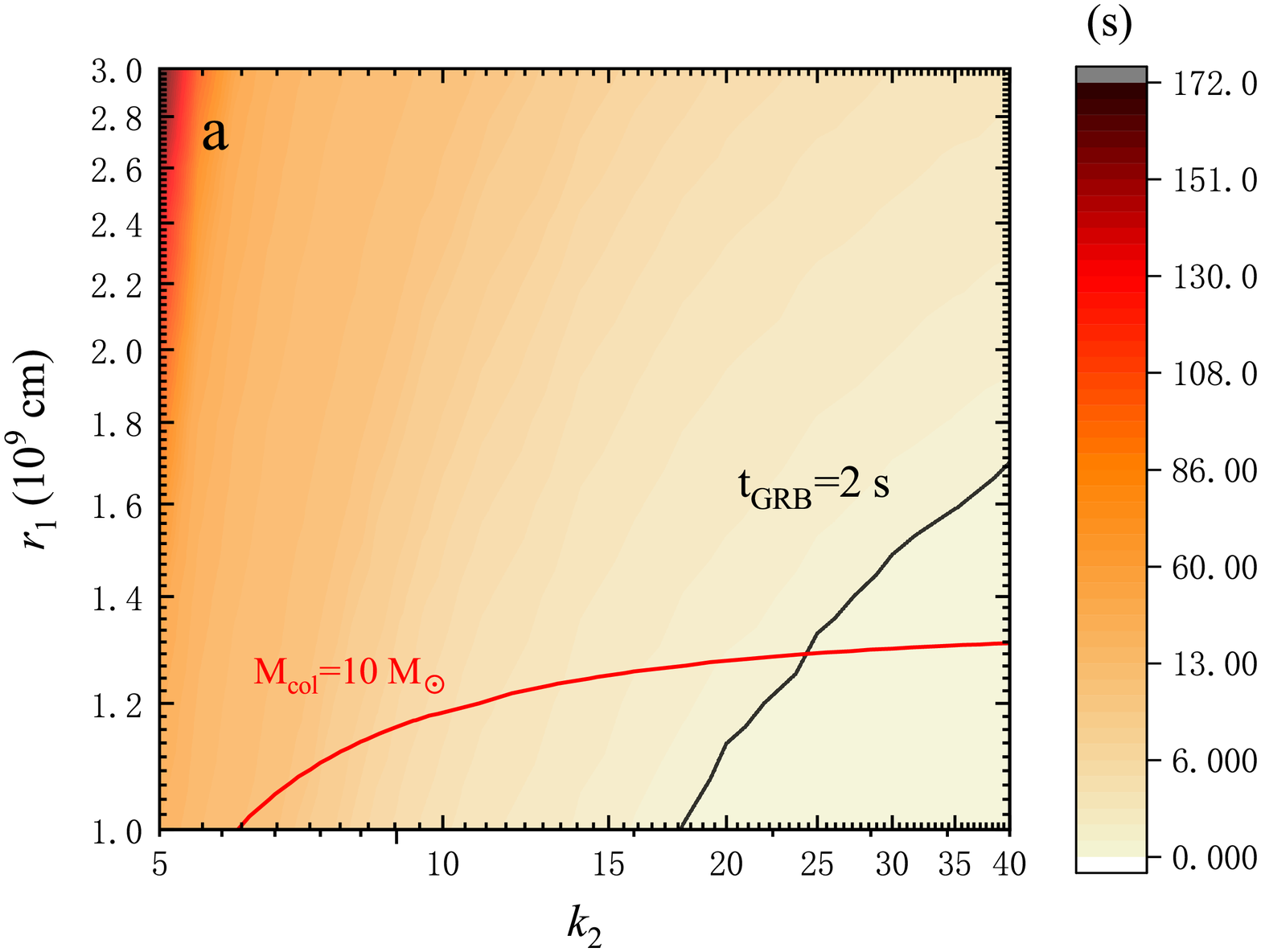}
\includegraphics[angle=0,scale=0.31]{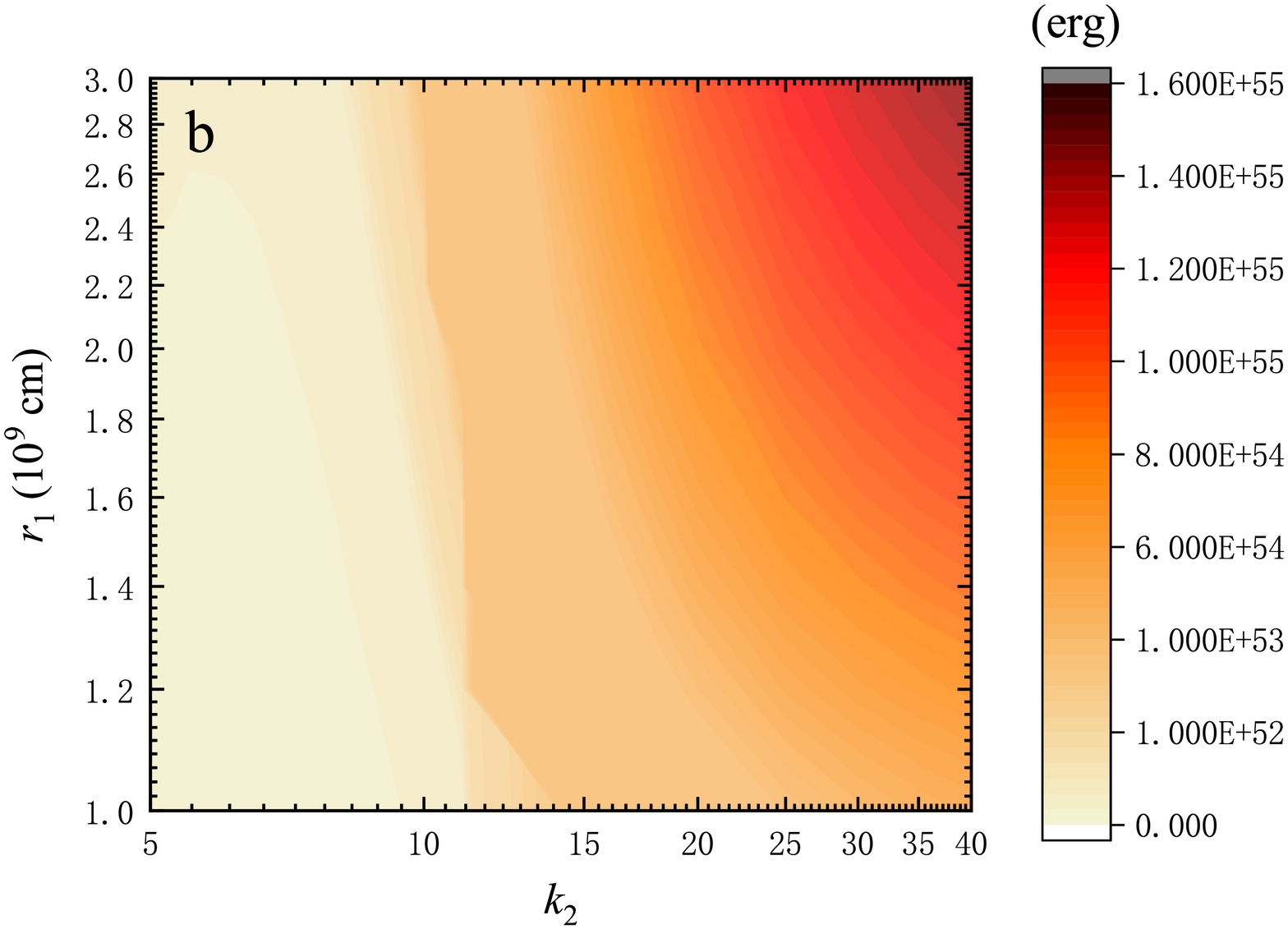}
\caption{Same as Figure 2 but for second jet model case.}
\end{figure*}

The GRB durations $t_{\rm{GRB}}$ and isotropic energy of prompt emission $E_{\rm{\gamma,iso}}$ of different collapsars for the first jet model are displayed in Figures 2(a) and 2(b), respectively. At the initial accretion stage, the mass supply to the BH hyperaccretion is from the core. The core becomes larger as $r_{1}$ becomes larger. Thus, a larger $r_{1}$ corresponds to a longer duration of the central engine. We assume that the central engine operates until the jet luminosity decreases to $10^{47}~\rm{erg}~\rm{s}^{-1}$. Therefore, the accretion of the envelope contributes to the duration of the central engine, especially when the envelope is thick. This contribution may be important in metal-free Pop III star. \citet{Suwa2011} investigated jet propagation in the envelope and showed that massive Pop III stars ($>100~M_{\odot}$) can produce GRBs even though they have large hydrogen envelopes due to the long-lasting accretion of the envelope itself. In addition, the jet propagation in light Pop III stars ($\sim 40~M_{\odot}$) has been investigated \citep[e.g.,][]{Nakauchi2012,Nagakura2012,Liu2018}. These studies found that light Pop III stars has the possibility of producing GRBs even though it keeps a massive hydrogen envelope. Some Pop III GRBs from high redshift might be detected as long-duration X-ray-rich GRBs by \emph{EXIST} or long-duration X-ray flashes by \emph{Lobster}. $k_{\rm{2}}$ determines the size and density of the envelope. There is competition between the mass supply onto the BH hyperaccretion and jet propagation into the envelope. Generally, a thick envelope can enhance the mass supply onto the BH hyperaccretion and increase the duration of the central engine, while a thin envelope allows the jet to break out of the collapsar quickly. When $k_{\rm{2}}$ is small, the collapsar has a thick envelope. Therefore, a smaller $k_{\rm{2}}$ corresponds to a longer duration of the central engine and a longer duration of jet propagation. In our work, the collapsar with a small $k_{\rm{2}}$ may come from a light Pop III star. When $k_{\rm{2}}$ is large, the collapsar almost loses its envelope, and both the duration of the central engine and the duration of jet propagation are short. As a result, collapsars with different density profiles can produce GRBs with different durations and $E_{\rm{\gamma,iso} }$ values.

The parameter space for GRBs with different durations is shown in Figure 2(a). The blank region corresponds to the case in which the jet fails to break out of the collapsar. The black line corresponds to $t_{\rm{GRB}}=2 ~\rm{s}$. Above this line, collapsars produce LGRBs. Obviously, a collapsar with a large core and a thin envelope is more likely to produce a LGRB. Below this line, collapsars would produce SGRBs or failed GRBs. When $k_{\rm{2}}$ is small, the jet can break out of the collapsar only when $r_{1}$ is large. This is because it takes a jet more time to break out of the collapsar when its envelope is thick. Therefore, the duration of the central engine should be long, i.e., $r_{1}$ should be large. We note that $t_{\rm{GRB}}=2 ~\rm{s}$ can be achieved regardless of the thickness of the envelope. However, these two situations are different. When $k_{\rm{2}}$ is small, a SGRB is mainly caused by the fact that the duration of the jet propagation is similar to the duration of the central engine. For a large $k_{\rm{2}}$, the jet can break out of the collapsar immediately, and an SGRB is caused by the fact that the duration of the central engine itself is short. We note that when $k_{\rm{2}}>10$, the envelope is already very thin and the jet can easily break out of the collapsar. $E_{\rm{\gamma,iso} }$ of different collapsars for the first model are shown in Figure 2(b). We can see that $E_{\rm{\gamma,iso} }$ increases as $k_{\rm{2}}$ increases. When the envelope is thick, it takes a long time to break out of the collapsar. Thus, a large part of the jet energy is consumed in the envelope, and the corresponding $E_{\rm{\gamma,iso} }$ is relatively small. As a result, GRBs from collapsars would have a specific duration and $E_{\rm{\gamma,iso} }$, which can help constrain the density profiles of collapsars.

The GRB duration $t_{\rm{GRB}}$ of different collapsars for the second jet model are displayed in Figure 3(a). We find that collapsars are more likely to produce LGRBs for the second jet model. SGRBs are produced only when $k_{\rm{2}}$ is large, i.e., the duration of the central engine itself is short. For the second jet model, even if the envelope is thick, the jet can easily beak out of the collapsar. Although a larger envelope may achieve $t_{\rm{eng}}\sim t_{b}$, the jet is nonrelativistic after breaking out of a very thick envelope. According to our result, the jet is nonrelativistic when $k_{\rm{2}}<5$. Therefore, for both jet models, collapsars are unlikely to produce SGRBs when $k_{\rm{2}}<5$.  Figure 3(b) shows $E_{\rm{\gamma,iso} }$ of different collapsars for the second model. For the same parameters, $E_{\rm{\gamma,iso} }$ in Figure 3(b) is larger than that in Figure 2(b). Similarly, $E_{\rm{\gamma,iso} }$ increases as $k_{\rm{2}}$ increases. For both jet models, collapsars with small $k_{\rm{2}}$ can produce GRBs. As a result, our calculations support the results of previous works \citep[e.g.,][]{Nakauchi2012,Nagakura2012} that light Pop III stars may produce GRBs. The Pop III GRBs from the early universe are expected to be detected in the future.

\section{Applications to GRB 200826A}

\begin{figure*}
\centering
\includegraphics[angle=0,scale=0.3]{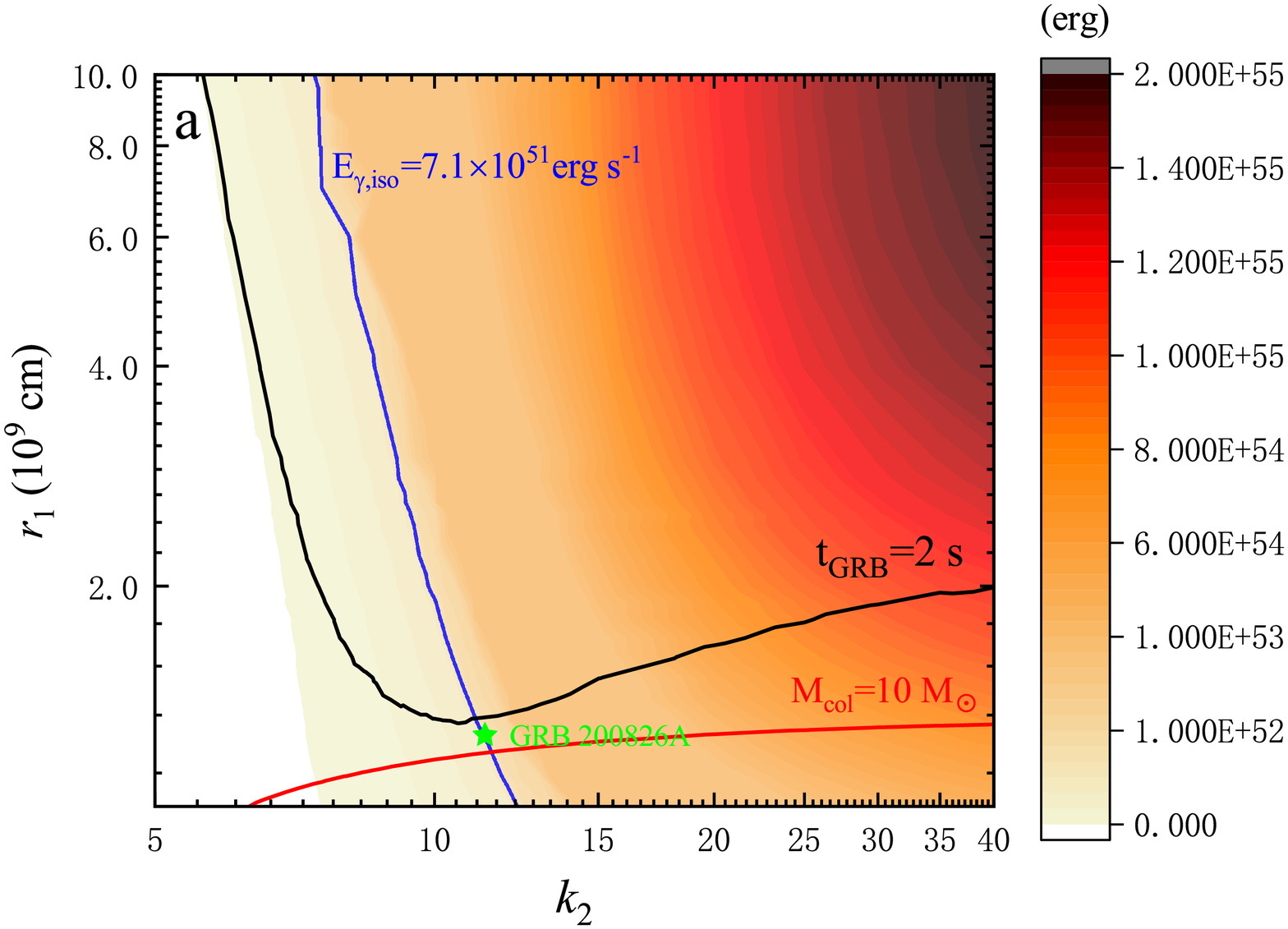}
\includegraphics[angle=0,scale=0.3]{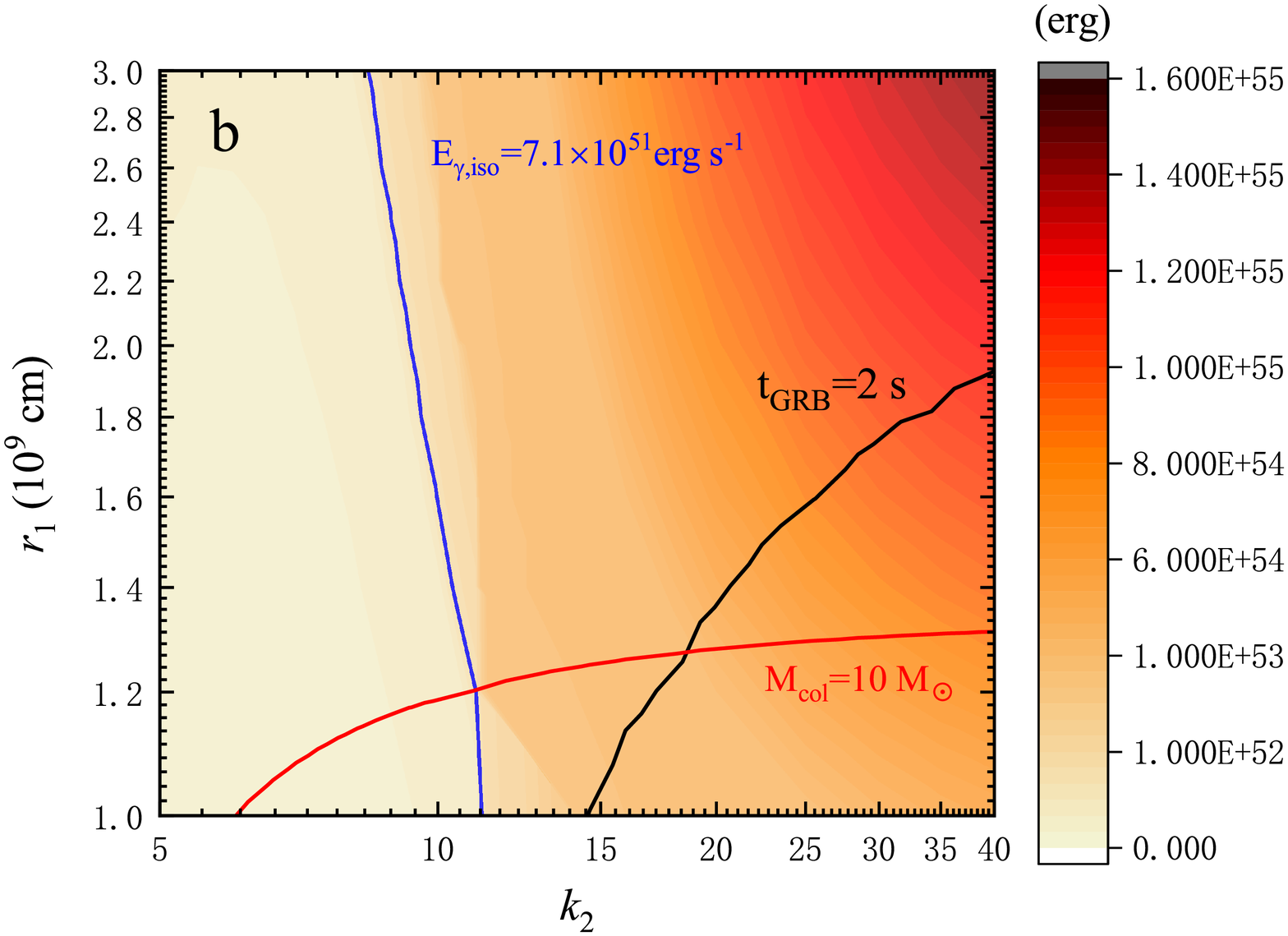}
\caption{Constrain collapsar density profiles of GRB200826A. Black and blue lines correspond to $t_{\rm{GRB}}=2 ~\rm{s}$ and $E_{\rm{\gamma,iso} }=7.1\times 10^{51}~\rm{erg}~\rm{s}^{-1}$, respectively. Red line corresponds to collapsar mass $M_{\rm col}=10 ~M_{\odot}$. Panels (a) and (b) show results of the first  and second jet models, respectively. Location of GRB 200826A is marked with green pentagram.}
\end{figure*}

GRB 200826A was detected by the Gamma-ray Burst Monitor (GBM) onboard the \emph{Fermi} Gamma-ray Space Telescope on August 26, 2020. The duration of this burst is $1.14\pm 0.13~\rm{s}$ in the $50-300$ keV energy range \citep{Ahumada2021}. \citet{Zhang2021} measured the ``amplitude parameter $f$" of this burst, which is defined as the ratio between the peak flux and the average background flux of the GRB lightcurve. They obtained $f=7.58\pm 1.23$ for GRB 200826A. According to \citet{Lu2014}, to make a LGRB a ``tip-of-iceberg" SGRB, the effective $f$ value is typically $<1.5$. The enormous $f$ value suggested that GRB 200826A is a genuine SGRB and cannot be the tip of iceberg of an LGRB. The optical afterglow of GRB 200826A was detected by the Large Binocular Telescope Observatory, and it helps identify the redshift $z \approx 0.7481$ \citep{Rothberg2020}. Therefore, the rest-frame duration of GRB 200826A should be $\sim 0.65~\rm{s}$, and the isotropic energy is $E_{\rm{\gamma,iso} }\sim7.1\times 10^{51}~\rm{erg}~\rm{s}^{-1}$. An supernova (SN) bump in the afterglow light curve was confirmed by \citet{Ahumada2020}. Meanwhile, additional optical observations of the afterglow and the SN have been reported \citep{Ahumada2021,Zhang2021,Rossi2022}. According to the spectral properties, host galaxy offset, total energy, and association with SN, GRB 200826A is considered to be the shortest LGRB from a collapsar. It appears to sit on the brink between a successful and a failed collapsar.

Here, we use our model to roughly constrain the collapsar density profile of GRB 200826A. According to our results, $t_{\rm{GRB}}<2 ~\rm{s}$ can be achieved by different density profiles. Therefore, we mainly constrain the density profile of GRB 200826A by isotropic energy, which is set as $E_{\rm{\gamma,iso} }=7.1\times 10^{51}~\rm{erg}~\rm{s}^{-1}$. First, we consider the first jet model, and the result is displayed in Figure 4(a). The black line corresponds to $t_{\rm{GRB}}=2 ~\rm{s}$, and the blue line corresponds to $E_{\rm{\gamma,iso} }=7.1\times 10^{51}~\rm{erg}~\rm{s}^{-1}$. The red line corresponds to collapsar mass $M_{\rm col}=10 ~M_{\odot}$. We mark the location of GRB 200826A in the figure with a green pentagram. The result shows that this burst might be produced by a collapsar whose envelope has been stripped. The mass of this collapsar is approximately $10~M_{\odot}$. Figure 4(b) displays the result for the second jet model. The black and blue lines correspond to $t_{\rm{GRB}}=2 ~\rm{s}$ and $E_{\rm{\gamma,iso} }=7.1\times 10^{51}~\rm{erg}~\rm{s}^{-1}$, respectively. We can see that the $E_{\rm{\gamma,iso} }$ of SGRBs from collapsars are generally higher than the $E_{\rm{\gamma,iso} }$ of GRB 200826A. Therefore, for the second jet model, the collapsar is unlikely to produce GRB 200826A even there is an efficiency of jet luminosity for the BZ mechanism or the outflows are considered. Moreover, SGRBs with relatively low $E_{\rm{\gamma,iso} }$ values are unlikely to be produced by collapsars for the second jet model. As a result, we suggest that GRB 200826A might come from a $\sim 10 ~M_{\odot}$ collapsar whose progenitor's envelope has been stripped and whose jets are driven by the annihilation of neutrinos.

\section{Summary}

In this paper, we investigated the propagation of jets in collapsars with different density profiles. We adopt two jet models. For the first jet model, the jet is driven by neutrino annihilation. For the second jet model, the jet is driven by the BZ process and neutrino annihilation together. For the same collapsar, the jet luminosity of the second model is obviously larger than that of the first model. There is competition between mass supply onto the BH hyperaccretion and jet propagation into the envelope. Although a thick envelope can increase the duration of the central engine, it also increases the duration of the jet propagation. As a result, the duration and $E_{\rm{\gamma,iso}}$ of GRBs from collapsars are determined by mass supply and jet propagation together.

We found that collapsars can produce LGRBs, SGRBs, and failed GRBs for both models. The density profiles for producing GRBs with different durations and $E_{\rm{\gamma,iso}}$ values are exhibited. Generally, a massive collapsar with a thin envelope is more likely to produce LGRBs. For the first jet model, both thick and thin envelopes can result in the production of SGRBs. For the second model, jets can easily break out of collapsars. We note that only collapsars with thin envelopes can give rise to SGRBs. Although a thick envelope can lead to $t_{\rm{eng}}\sim t_{\rm b}$, the jet would be nonrelativistic after it breaks out of the collapsar. Thus, for the second model, collapsars are more likely to produce LGRBs. The thickness of the envelope can significantly affect $E_{\rm{\gamma,iso}}$. $E_{\rm{\gamma,iso}}$ increases as the envelope becomes thinner.

GRB 200826A is considered to be an SGRB from a collapsar. We show that this burst might be produced by a $\sim 10~M_{\odot}$ collapsar whose envelope has been stripped, and the jets should be launched by the neutrino annihilation process.

Note that the propagation of the jet is very complicated. Many factors, such as the jet opening angle, star rotation, circumstellar medium, jet feedback, and disk outflow feedback \citep{Liu2019} affect the propagation of the jet and GRB production. Therefore, it is inadequate to constrain the properties of collapsars solely according to the duration and $E_{\rm{\gamma,iso}}$ of GRBs. The joint multimessenger observations, i.e., MeV neutrinos and gravitational waves (GWs), can help investigate GRB physics. In \citet{Wei2019} and \citet{Wei2020}, we proposed that whether the jets are chocked in the envelopes or not, the activities of the central compact objects will produce the detectable MeV neutrinos and GWs. Furthermore, we presented that the strong GW signals from collapsars, jets, and central engines in the Local Group can be detected by the operational or planned GW detectors \citep[e.g.,][]{Wei2020}.

\acknowledgments
This work was supported by the National Natural Science Foundation of China under grant 12173031.

\end{document}